\documentclass[12pt]{iopart}

\usepackage{amstext}
\usepackage{graphics}
\usepackage{graphicx}
\usepackage[english]{babel}
\usepackage[numbers,sort&compress]{natbib}	
\newcommand{\beq}{\begin{eqnarray}}
\newcommand{\eeq}{\end{eqnarray}}
\newcommand{\be}{\begin{equation}}
\newcommand{\ee}{\end{equation}}

\begin{document}

\def\aligned{\vcenter\bgroup\let\\\cr
\halign\bgroup&\hfil${}##{}$&${}##{}$\hfil\cr}
\def\endaligned{\crcr\egroup\egroup}

\pagestyle{plain}
\title[Multi-terminal multi-junction dc SQUID]{Multi-terminal multi-junction dc SQUID for nanoscale magnetometry}
\author{Alexander Y. Meltzer, Aviram Uri and Eli Zeldov}
\address{Department of Condensed Matter Physics, Weizmann Institute of Science, Rehovot 76100, Israel}\
\eads{\mailto{alexander.meltzer@weizmann.ac.il}, \mailto{aviram.uri@weizmann.ac.il}, \mailto{eli.zeldov@weizmann.ac.il}}

\begin{abstract}
Miniaturization of superconducting quantum interference devices (SQUIDs) is of major importance for the development of sensitive scanning nanoscale magnetometry tools. The high sensitivity of nanoSQUIDs is restricted, however, to only particular periodic values of the applied magnetic field, making accurate measurements at intermediate values of the field impossible. We present a theoretical investigation of a multi-terminal, multi-junction SQUID (mSQUID) that lifts this limitation by providing electrical means for a continuous shift of the quantum interference pattern with respect to the applied field. Analysis of 4-terminal, 4-junction and 3-terminal, 3-junction mSQUIDs shows that operation at maximum sensitivity can be obtained at any value of the magnetic field by applying control current to the extra terminals. The model describes the variation and the shift of the interference pattern as a function of the control currents, junction asymmetries, and the mSQUID inductance. The mSQUID is also shown to provide a direct measurement of the current-phase relations of superconducting junctions. The derived model provides a quantitative description of the recently developed multi-terminal nanoSQUID-on-tip.
\end{abstract}

\noindent{\it Keywords\/}: multi-terminal SQUID, SQUID-on-tip, magnetometry, current-phase relations\\

\submitto{\SUST}
\maketitle
\section{Introduction}

Superconducting quantum interference devices (SQUIDs) are very sensitive sensors of magnetic field \cite{NANO1,NANO2,NANO3,NANO5,NANO6,NANO7,Schwarz2015,Hazra2013,Hao2008,Clarke2004,Cleuziou2006} and in recent years are widely used for nanoscale magnetic sensing and for scanning magnetic microscopy \cite{SCANNING1,SCANNING2,SCANNING3,SCANNING4,SCANNING5,SCANNING6,SCANNING7,SCANNING8,SCANNING9,Kirtley2010,comp1,comp2,Lachman2015,Kremen2016,Granata2015,Hykel2014}. Scanning SQUIDs are commonly fabricated using planar lithographic techniques and often include integrated pickup and feedback coils, which allow flux biasing the SQUID near its optimal working point using a flux-locked loop (FLL) \cite{Clarke2004}. Since the SQUID and the pickup coil are separated in space, the SQUID can be maintained at its optimal flux bias conditions, while the
measured local magnetic field of the sample may vary substantially.

Recently, a new technique for fabrication of nanoSQUIDs has been introduced, in which the SQUID loop is fabricated on the apex of a sharp pipette using self-aligned deposition method \cite{SOT1, SOT2, SOT3, SOT4, SOT5}. These SQUID-on-tip (SOT) sensors are highly advantageous for scanning SQUID microscopy due to their very small size, close proximity to the sample surface, high spin sensitivity, and operation in high fields. The geometry of the SOT, however, does not allow integration of a feedback loop and the nanoscale proximity to the sample dictates that the flux in the SQUID loop cannot be adjusted independently from the local magnetic field of the sample. As a result the sensitivity of the device depends on the value of the magnetic field applied to the sample. More specifically, the critical current of a SQUID, $I_c(\Phi_a)$, is periodic in the externally applied flux
$\Phi_a$ with a period of flux quantum $\Phi_0$. The sensitivity of the SQUID is roughly proportional to the flux-to-current transfer  function $|dI_c/d\Phi_a|$, which is also periodic. As a result, the SOT is most sensitive around the points where the transfer  function has its maxima and is markedly less sensitive at all other values of the magnetic field resulting in blind spots. This is a significant drawback that limits the performance of the SOT.

In order to overcome this limitation of the SOT and allow accurate measurements over an extended field range, we introduce a multi-junction, multi-terminal SQUID (mSQUID). In contrast to regular two-junction SQUIDs, we show that the critical current interference pattern in mSQUIDs can be shifted continuously with respect to the applied flux $\Phi_a$ by applying control currents to the additional terminals, as has been recently demonstrated experimentally \cite{ExperPaper}. The resulting electrically controllable interference pattern shift allows the SQUID to operate at maximum sensitivity over its full range of operational magnetic field.

In this paper, we derive a mathematical model for description of a 4-junction, 4-terminal mSQUID with two control currents and a 3-junction, 3-terminal mSQUID with one control current (see Fig. \ref{fig1}). This model enables us to analyze the influence of control currents, junction asymmetry and self-inductance on the shape of the interference pattern and on its shift. In addition,  the modulation depth of the critical current and the skewness of the interference pattern are discussed.

Four-junction SQUID configurations were previously discussed theoretically in the context of control of the critical current \cite{Fink1987} and of current amplification in micronetworks \cite{Larson1995}; however, the shift of the interference pattern and the problem of blind spots has not been addressed. Several other multi-junction or multi-terminal configurations, which provide various functionalities and flux bias of the SQUID, have been studied in the past \cite{SOT5,Schwarz2015,Chiorescu2003,MZB,3jj1,3jj2,3jj3,3jj4}. The unique feature of the described mSQUID is the possibility of external control of the superconducting phase difference across the individual junctions without substantially affecting the flux in the SQUID loop.

The structure of this paper is as follows: in Section 2 we derive the stationary analytical model of the 4-terminal, 4-junction mSQUID and analyze its interference pattern, in Section 3 we present the derivation and analysis of a model for the 3-junction, 3-terminal mSQUID, and a brief summary in Section 4 concludes the paper.

\section{Four-terminal mSQUID \label{modelSec}}
\subsection{The general model \label{ss1}}
\begin{figure}[c]
  \makebox[\textwidth]{\includegraphics[trim=0 0 0 0, clip,width=.7\textwidth]{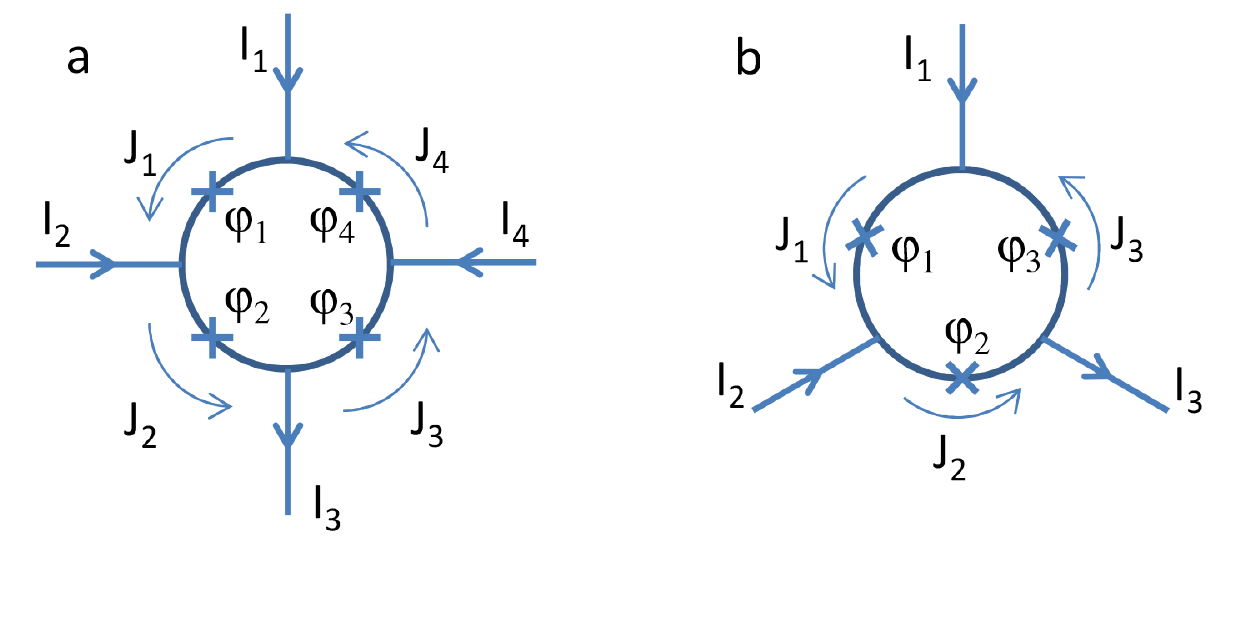}}
  \caption{Schematic layout of 4-junction (a) and 3-junction (b) mSQUID. $I_1$ is the bias current, $I_2$ and $I_4$ in (a) and $I_2$ in (b) are the control currents, and $I_3$ is the drain current. The currents through the junctions are described by $J_{k}=a_kI_0\sin(\varphi_k)$.}
  \label{fig1}
\end{figure}

We consider an mSQUID consisting of four Josephson junctions (Fig. \ref{fig1}) operating under stationary conditions. The critical currents are given by $a_kI_0$, $k=1,2,3,4$, where $a_k$ is the asymmetry coefficient for the $k$th junction. The values of $a_k$ are not restricted to a certain range and depend on the choice of $I_0$. 
We assume that the junctions follow the standard sinusoidal current-phase relations $J_{k}=a_kI_0\sin(\varphi_k)$, where $\varphi_k$ is the phase difference across the $k$th junction \cite{Likharev1979}. The relation between the net magnetic flux $\Phi_{tot}$ in the mSQUID loop and the phase differences $\varphi_k$ across the four junctions is given by
\be
\varphi_1+\varphi_2+\varphi_3+\varphi_4 + 2\pi \frac{\Phi_{tot}}{\Phi_0}=2\pi n, \label{vvvv}
\ee
where $n$ is an integer.

All the phase differences and the currents are  oriented in a counterclockwise direction, as depicted in Fig. \ref{fig1}. The total flux $\Phi_{tot}$ is the sum of the applied flux $\Phi_a$ and the self-induced flux,
\be
\Phi_{tot}=\Phi_a + L_1J_1+L_2J_2+L_3J_3+L_4J_4,
\ee
where $L_k$ is the geometric inductance of each segment. In the symmetric case, we have $L_k=L/4$, where $L$ is the loop inductance. The junction currents $J_k$ and the external currents $I_k$ are related by the Kirchhoff law as follows:
 \be
I_1 = J_1-J_4, \quad I_2 = J_2-J_1, \quad I_3 = J_2-J_3, \quad I_4 = J_4-J_3. \label{oneof}
 \ee
Here, $I_1$ is the bias current, $I_2$ and $I_4$ are the control currents, and $I_3$ is the drain current.

In order to work with dimensionless quantities, we normalize the currents to $I_0$ and the flux to $\Phi_0$ and denote the normalized variables using lower case letters. In the new variables, we can rewrite (\ref{vvvv}) as
\be
\varphi_1+\varphi_2+\varphi_3+\varphi_4 + 2\pi \phi_a+\pi\beta_L j=2\pi n, \label{vvvv2}
\ee
where $\beta_L=2I_0L/\Phi_0$ and $j=(j_1+j_2+j_3+j_4)/4$ is the circulating current in the mSQUID.

The relation between the phase differences $\varphi_j$ when the mSQUID is in a critical state can be derived using the method of Lagrange multipliers \cite{TVD}. Using (\ref{vvvv2}) and the expressions for $i_2$ and $i_4$ in (\ref{oneof}) as side conditions, we write the Lagrangian
\be
\begin{aligned}
\mathcal{L}(\varphi_1,\varphi_2,\varphi_3,\varphi_4,\lambda_1,\lambda_2,\lambda_3) = & a_1\sin(\varphi_1)-a_4\sin(\varphi_4)\\&+ \lambda_1(\varphi_1+\varphi_2+\varphi_3+\varphi_4 +\pi\beta_L j)\\& + \lambda_2(i_4-a_4\sin(\varphi_4)+a_3\sin(\varphi_3))\\&  + \lambda_3(i_2-a_2\sin(\varphi_2)+a_1\sin(\varphi_1)).
\end{aligned}
\ee
The  problem of finding the  critical current through terminal 1, $i_{c1}$, of the mSQUID is equivalent to that of finding the critical points  of $\mathcal{L}$. We now proceed by taking the derivatives of $\mathcal{L}$ with respect to each independent variable, equating each of these expressions to zero and eliminating the Lagrange multipliers $\lambda_m$.

Let $\varphi_{ck}$ denote the critical-state phase differences of the mSQUID. The extreme solution, found by the method of Lagrange multipliers is given by
\be
a_4\cos\varphi_{c4}+ g+\frac{a_4\cos\varphi_{c4}}{a_3\cos\varphi_{c3}}g=0, \label{e2}
\ee
where
\be
g = \frac{\cos\varphi_{c1} \cos\varphi_{c2}}{r\cos\varphi_{c1}+\frac1 r\cos\varphi_{c2}+\pi\beta_L \cos\varphi_{c1} \cos\varphi_{c2}} \label{e3}
\ee
and $r=a_1/a_2$. In order to solve (\ref{e2})-(\ref{e3}) we start by assigning values to $\varphi_{c1}$ and $\varphi_{c2}$ on a two-dimensional grid. Next, using (\ref{e3}) we find $g$ and proceed to solve for (\ref{e2}), which in our setup has only one unknown because the two phases $\varphi_{c3}$ and $\varphi_{c4}$ are related by
\be
i_4=a_4\sin\varphi_{c4}-a_3\sin\varphi_{c3}.
\ee
Using Equation (\theequation) and assuming $-\pi<\varphi_{c4}<\pi$, we find
\beq
a_{3}\cos\varphi_{c3} &=&\left\{\begin{array}{ll}
                                   \sqrt{a_3^2-(-i_4+a_4\sin\varphi_{c4})^2}, & -\frac{\pi}2\leq \varphi_{c3}\leq \frac{\pi}2, \\
                                   -\sqrt{a_3^2-(-i_4+a_4\sin\varphi_{c4})^2}, & \text{otherwise},
                                 \end{array}
 \right.\nonumber\\\label{e155}
\eeq
and
\beq
\varphi_{c3} &=&\left\{\begin{array}{ll}
                                   \arcsin((a_4\sin(\varphi_{c4})-i_4)/a_3), & -\frac{\pi}2\leq \varphi_{c3}\leq \frac{\pi}2, \\
                                   \pi-\arcsin((a_4\sin(\varphi_{c4})-i_4)/a_3), & \frac{\pi}2< \varphi_{c3}\leq \pi, \\
                                   -\pi-\arcsin((a_4\sin(\varphi_{c4})-i_4)/a_3), & \text{otherwise}.
                                 \end{array}
 \right.\nonumber\\ \label{e15}
\eeq
Since the value of $\varphi_{c3}$ is not known, we use all possible combinations defined by  (\ref{e155})-(\ref{e15}) for the solution.

The above model, consisting of (\ref{vvvv2}) and (\ref{e2})-(\ref{e15}), allows the numerical calculation of all the phase differences $\varphi_{ck}$, the corresponding critical current $i_{c1}$, and the applied flux $\phi_a$ for various values of the coefficients $a_k$, $\beta_L$ and the control currents $i_2$ and $i_4$. Since the model is invariant under the simultaneous transformation  $\varphi_{ck}\rightarrow - \varphi_{ck}$ and $\phi_a\rightarrow - \phi_a$, the negative critical current can be found from the positive critical current by transforming $i_k\rightarrow - i_k$ and $\phi_a\rightarrow - \phi_a$.

\begin{figure}[!htb]
  \makebox[\textwidth]{\hspace*{5cm}\includegraphics[trim=50 350 50 50, clip,scale=1, width=\textwidth]{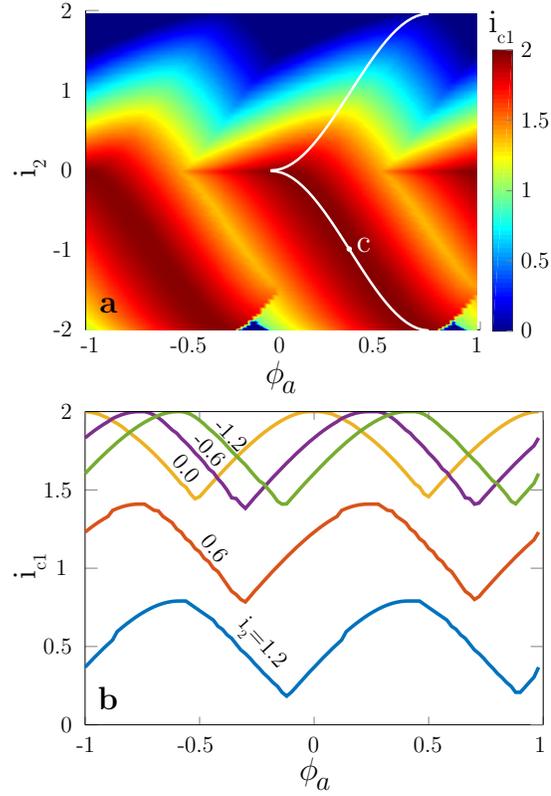}}
  \caption{(a) The critical current $i_{c1}(\phi_a,i_2)$ of the symmetric 4-junction mSQUID with $a_k=1$, $i_4=0$ and $\beta_L=1$ as a function of the control current $i_2$ and the applied flux $\phi_a$. The white line $\phi_{a}^{max}(i_2)$ traces the location of the maximum critical current $i_{c1}^{max}$. Its lower branch reflects the current-phase relation of junction 2 and the upper branch the current-phase relation of junction 1. Point (c) corresponds to $j_2=0$. (b) Interference pattern $i_{c1} (\phi_a)$ (horizontal line cuts through (a)) for different values of the control current $i_2 = -1.2,\,-0.6,\,0,\,0.6$, and $1.2$  showing the controllable shift of the patterns by $i_2$.}
  \label{Fig2}
\end{figure}

\begin{figure}[!htb]
  \makebox[\textwidth]{\hspace*{5cm}\includegraphics[trim=150 370 50 250, clip,scale=1, width=\textwidth]{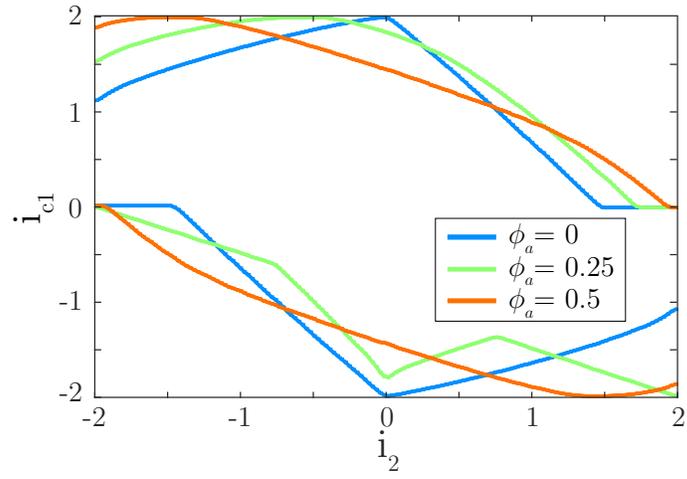}}
  \caption{Dependence of the critical current $i_{c1} (i_2)$ on the control current $i_2$ for different values of the applied flux $\phi_a=0,\,0.25,\,0.5$ (vertical line cuts through Fig \ref{Fig2}(a)). Both positive and negative values of $i_{c1}$ are shown.}
  \label{Fig25}
\end{figure}

A typical form of the critical current surface $i_{c1}(\phi_a,i_2)$ calculated using our model is shown in Fig. \ref{Fig2}(a) for the case of $i_4=0$, $a_k=1$ and $\beta_L=1$. Interference patterns for several horizontal line cuts through Fig. \ref{Fig2}(a) are shown in Fig. \ref{Fig2}(b). Evidently the control current $i_2$ shifts the interference pattern and, consequently, the location of the optimal working points and of the low sensitivity areas of the mSQUID. Figure \ref{Fig25} shows several vertical line cuts through Fig. \ref{Fig2}(a), illustrating the dependence of $i_{c1}$ on $i_2$ at selected values of the applied flux. Figure \ref{Fig25} also shows the corresponding negative values of $i_{c1}$ when the mSQUID is biased by a negative $i_{1}$. For each $\phi_a$, the central area between the positive and negative $i_{c1}$ curves defines the phase space for which the mSQUID is in superconducting state with $|i_1|<|i_{c1}|$ and hence has no field sensitivity.

\subsection{Shift of the mSQUID interference pattern \label{TOTMP}}

We now provide a detailed analysis of the effect of the control currents on the quantum interference patterns. The surface $i_{c1}(\phi_a,i_2)$ in Fig. \ref{Fig2}(a) has a well-specified structure and is divided into two parts by a horizontal demarcation line $i_2=a_2-a_1$, which for a symmetric mSQUID is given by $i_2=0$. For $i_2<a_2-a_1$ the entire interference pattern $i_{c1}(\phi_a)$ is continuously shifted to the left along $\phi_a$ axis as $i_2$ increases, while for $i_2>a_2-a_1$, in contrast, the interference pattern is shifted to the right. The current $i_2$ itself is bound by $-(a_2+a_1)\leq i_2\leq a_2+a_1$.

It is beneficial to trace the location of the maximum of the critical current $i_{c1}^{max}(i_2)$ on the surface $i_{c1}(\phi_a,i_2)$ (for a fixed value of $i_4$) as shown by the white line in Fig. \ref{Fig2}(a). This line, denoted $\phi_{a}^{max}(i_2)$, is of particular value since it follows a simple mathematical description and provides a clear insight into the underlying mechanism of the interference shift. In particular, as shown below, the following conditions hold along the $\phi_{a}^{max}(i_2)$ line: i) The right and left arms of the mSQUID behave independently. ii) The current in the right arm is determined only by the values of $a_3$, $a_4$, and $i_4$ and is constant for a fixed $i_4$. iii) The current in the left arm is determined only by the values of $a_1$, $a_2$, and $i_2$ and therefore is controllable by $i_2$. iv) The control current $i_2$ determines which of the junctions 1 or 2 is in the critical state with a phase drop of $\pi/2$. v) For $i_2\leq a_2-a_1$ the phase $\varphi_{1}$ equals $\varphi_{c1}=\pi/2$ independently of $i_2$ and hence $i_{c1}^{max}$ is constant. The value of $\varphi_{c2}$, in contrast, increases with $i_2$ resulting in a controllable shift of the interference pattern to the left. vi) For $i_2>a_2-a_1$ the phase $\varphi_{2}$ equals $\varphi_{c2}=\pi/2$ independently of $i_2$ while $\varphi_{c1}$ decreases with increasing $i_2$. As a result, increasing $i_2$ causes a decrease in $i_{c1}^{max}$ and a controllable shift of the interference pattern to the right.

The above listed properties of the mSQUID along the curve $\phi_{a}^{max}(i_2)$  are derived as follows. We first note that in the critical state  the phase differences satisfy
\beq
\begin{array}{ll}
\varphi_{c1}=\pi/2,& \text{for}\;i_2\leq a_2-a_1, \\
\varphi_{c2}=\pi/2,& \text{for}\;i_2>a_2-a_1,\\
\varphi_{c3}=-\pi/2,& \text{for}\;i_4>a_3 -a_4,\\
\varphi_{c4}=-\pi/2,& \text{for}\;i_4\leq a_3 -a_4.
\end{array}\label{car1}
\eeq
From (\ref{car1}) and (\ref{oneof}), we find that
\beq
\begin{array}{ll}
j_3&=-\min(a_3,a_4+i_4),\\
j_4&=-\min(a_3-i_4,a_4).
\end{array}\label{j34}
\eeq
Summing the two first expressions in (\ref{oneof}) we find
\be
i_{c1}=j_2-j_4-i_2. \label{ic1}
\ee
Inserting the expression for $j_4$ given in (\ref{j34}) into (\ref{ic1}) and using (\ref{car1}), we obtain the following equation describing $i_{c1}^{max}$ above and below the demarcation line:
\be
i_{c1}^{max}=\left\{ \begin{array}{ll}
                                                a_2+\min(a_3-i_4,a_4)-i_2, & i_2> a_2-a_1, \\
                                                a_1+\min(a_3-i_4,a_4), & i_2\leq a_2-a_1.
                                              \end{array}
\right.\label{ic11}
\ee
From (\ref{ic11}), we deduce that for a fixed value of $i_4$, the value of $i_{c1}^{max}(i_2)$ remains constant below the demarcation line and decreases linearly with $i_2$ above it, as seen in Fig. \ref{Fig2}(a). Since for $i_2\leq a_2-a_1$ the value of $i_{c1}^{max}$ remains constant and $i_2$ controls only the shift of the interference pattern, operation below the demarcation line is more favorable for practical applications of the mSQUID.

We now analyze the shift of the interference pattern that is described by $\phi_{a}^{max}(i_2)$ curve. Along this curve $j_3$ and $j_4$ are constant for a fixed $i_4$. Using (\ref{vvvv2}), (\ref{car1}) and (\ref{j34}), we deduce that for $i_2\leq a_2-a_1$ the curve $\phi_{a}^{max}(i_2)$ is described by the relation
\be
\begin{aligned}
\varphi_{c2}=&-\pi/2-\arcsin(j_3/a_3)-\arcsin(j_4/a_4)\\&-2\pi\phi_{a}^{max}-\pi\beta_L(a_1+a_2\sin\varphi_{c2}+j_3+j_4)/4, \end{aligned}
\label{dd}
\ee
where the phase difference $\varphi_{c2}$ is determined by $i_2=a_2\sin \varphi_{c2}-a_1$. Hence, below the demarcation line the current $i_2$ controls the interference pattern shift by electrically controlling the superconducting phase difference $\varphi_{c2}$.

Similarly, we can deduce that for $i_2>a_2-a_1$ the curve $\phi_{a}^{max}(i_2)$ satisfies the relation
\be
\begin{aligned}
\varphi_{c1}=&-\pi/2-\arcsin(j_3/a_3)-\arcsin(j_4/a_4)\\&-2\pi\phi_a^{max}-\pi\beta_L(a_1\sin\varphi_{c1}+a_2+j_3+j_4)/4,
\end{aligned}
\label{dd2}
\ee
with phase difference $\varphi_{c1}$ determined by the identity $i_2=a_2-a_1\sin \varphi_{c1}$. Thus above the demarcation line $i_2$ provides an electrical control of the superconducting phase difference $\varphi_{c1}$.

\begin{figure}[!htb]
\centering
  \makebox[\textwidth]{\includegraphics[trim=50 270 0 150, clip,scale=.9]{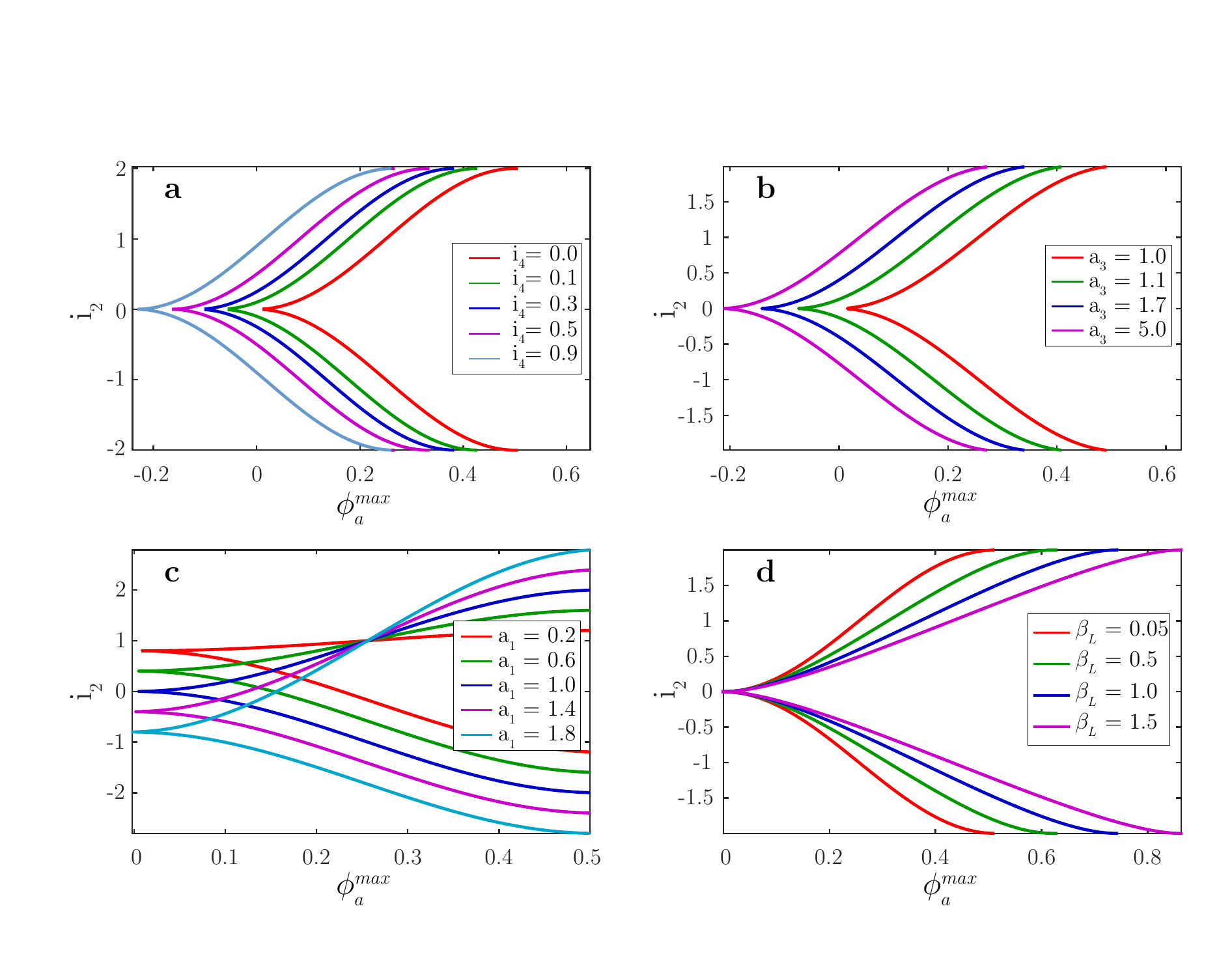}}
  \caption{The function $\phi_{a}^{max}(i_2)$ for the 4-terminal mSQUID,which describes the shift of the interference pattern by control current $i_2$, for the following parameter values: (a) $a_k=1$, $\beta_L=0.05$, and $i_4=0,\,0.1,\,0.3,\,0.5,$ $0.9$. (b) $a_1=1$, $a_2=1$, $a_4=1$, $\beta_L=0.05$, $i_4=0$, and $a_3=1,\,1.1,\,1.7,$ $5$. (c)  $a_2=1$, $a_3=1$, $a_4=1$, $\beta_L=0.05$, $i_4=0$, and $a_1=0.2,\,0.6,\,1,\,1.4,$ $1.8$. (d) $a_k=1$, $i_4=0$, and $\beta_L=0.05,\,0.5,\,1,$ $1.5$.}
  \label{fig3}
\end{figure}

Figure \ref{fig3} illustrates the dependence of the curve $\phi_{a}^{max}(i_2)$ on various parameters. The currents $j_3$ and $j_4$ in (\ref{dd}) and (\ref{dd2}) are uniquely defined by $i_4$, as shown in (\ref{j34}), and are constant for a fixed $i_4$. Varying $i_4$ therefore does not change the shape of $\phi_{a}^{max}(i_2)$ and only displaces it horizontally as shown in Fig. \ref{fig3}(a). For the same reason, changing $a_3$ or $a_4$ results only in a horizontal displacement. In Fig. \ref{fig3}(b), this result is plotted for various values of $a_3$. Varying $a_1$ or $a_2$, in contrast, does change the shape of $\phi_{a}^{max}(i_2)$, but not in a trivial way. Since the demarcation line is given by $a_2-a_1$, variation in $a_1$ or $a_2$ results in a vertical shift of $\phi_{a}^{max}(i_2)$ as seen in Fig. \ref{fig3}(c) for the case of varying $a_1$. In addition, the vertical extent of $\phi_{a}^{max}(i_2)$ below the demarcation line is given by $2a_2$, while above the line it equals $2a_1$. Reducing $a_1$ thus shrinks the upper branch of $\phi_{a}^{max}(i_2)$ while keeping its lower branch intact, as shown in Fig. \ref{fig3}(c). Varying $a_2$ will have an opposite effect.

The dependence of $\phi_{a}^{max}$ on various parameters in Fig. \ref{fig3} can be further analyzed using the following transformations which leave (\ref{dd})-(\ref{dd2}) invariant.  It can be readily shown that, for any $\beta_L$,  (\ref{dd})-(\ref{dd2}) are invariant under the following transformations:
\beq
&G_1:& \quad i_2\rightarrow -i_2,\quad a_1\leftrightarrow a_2, \quad \varphi_{c1}\leftrightarrow\varphi_{c2},\label{trtr1}\\
&G_2:& \quad i_4\rightarrow -i_4,\quad a_3\leftrightarrow a_4.\label{trtr2}
\eeq
Transformation $G_1$ consists of changing the sign of $i_2$ and the replacement of $a_1$ with $a_2$ and of $\varphi_{c1}$ with $\varphi_{c2}$. The replacement of phase $\varphi_{c3}$ with $\varphi_{c4}$ in $G_2$ is implied but not  explicitly stated because these phase differences are not present in (\ref{dd})-(\ref{dd2}) . For $\beta_L=0$ we can write two additional transformations, which keep (\ref{dd})-(\ref{dd2}) invariant:
\beq
&G_3:& \quad a_3\rightarrow 1/a_3, \quad a_4\rightarrow 1/a_4,\quad i_4\rightarrow -\frac{i_4}{a_3 a_4}, \label{trtr3}\\
&G_4:& \quad a_3\rightarrow 1/a_4, \quad a_4\rightarrow 1/a_3,\quad i_4\rightarrow  \frac{i_4}{a_3 a_4}. \label{trtr4}
\eeq
We note that the composite tranformation $G_3\circ G_4=G_2$ for $\beta_L=0$.

Under transformation $G_1$, we find that by varying $a_2$ instead of $a_1$ we obtain the reflection of the curves in Fig. \ref{fig3}(c) about the x-axis. We also find that any decrease in $i_4$ when $i_4<0$ under transformation $G_2$ results in a shift of $\phi_{a}^{max}(i_2)$ to the left, similar to the effect of increasing $i_4$ when $i_4>0$ in Fig. \ref{fig3}(a). In Fig. \ref{fig3}(b) $i_4=0$ and $a_4=1$, therefore, under transformation $G_3$, a decrease in $a_3$ for $a_3<1$ shifts the curve $\phi_{a}^{max}(i_2)$ to the left, similarly to an increase in $a_3$ for $a_3>1$.

We now discuss the horizontal extent of $\phi_{a}^{max}(i_2)$. This is a particularly important parameter since it defines the maximum possible shift of the interference pattern. In regular two-junction SQUIDs, the most sensitive working points are found at flux values of $\phi_a\simeq 1/4+n/2$. In order to have a sensitive response at any value of the applied flux, the ability to shift the interference pattern electrically by at least $1/2$ is therefore required. This means that the horizontal extent of $\phi_{a}^{max}(i_2)$ should be at least $1/2$. Since the critical-state phase differences satisfy $\varphi_{c1}=-\pi/2$ for $i_2=a_2+a_1$, $\varphi_{c1}=\varphi_{c2}=\pi/2$ for $i_2=a_2-a_1$ and $\varphi_{c2}=-\pi/2$ for $i_2=-(a_2+a_1)$ we can find, using (\ref{dd}) and (\ref{dd2}), that the horizontal extent of $\phi_{a}^{max}(i_2)$ is $1/2+\beta_L a_1/4$ when $i_2>a_2-a_1$ and $1/2+\beta_L a_2/4$ when $i_2\leq a_2-a_1$. As shown in Fig. \ref{fig3}, the horizontal span of $\phi_{a}^{max}(i_2)$ is indeed $1/2$ for $\beta_L\ll1$ and greater then $1/2$ for larger $\beta_L$, thus giving the mSQUID the novel ability of highly sensitive operation over its entire range of operating fields. This powerful property has been recently demonstrated experimentally in multi-terminal SOT \cite{ExperPaper}.

Besides controlling the optimal working point, the electrical tunability of the mSQUID can be utilized for noise reduction. Some common noise reduction schemes \cite{Clarke2004} are based on periodic flux-bias switching of the SQUIDs, which in the case of mSQUID can be readily achieved electrically. These schemes, however, may require flux bias switching by up to a full period of $\Phi_0$. This requirement can be attained in mSQUID by extending the span of $\phi_{a}^{max}(i_2)$ by either of the following two methods. Figure \ref{fig3}(d) shows that for $\beta_L<<1$ the span of $\phi_{a}^{max}(i_2)$ is $1/2$, but it increases substantially upon increasing $\beta_L$, reaching $0.75$ at $\beta_L=1$. Increasing $\beta_L$ much further is undesirable because of the accompanying reduction in the modulation depth.

\begin{figure}[!htb]
\centering
  \makebox[\textwidth]{\hspace*{10cm}\includegraphics[trim=50 200 50 150, clip,scale=1]{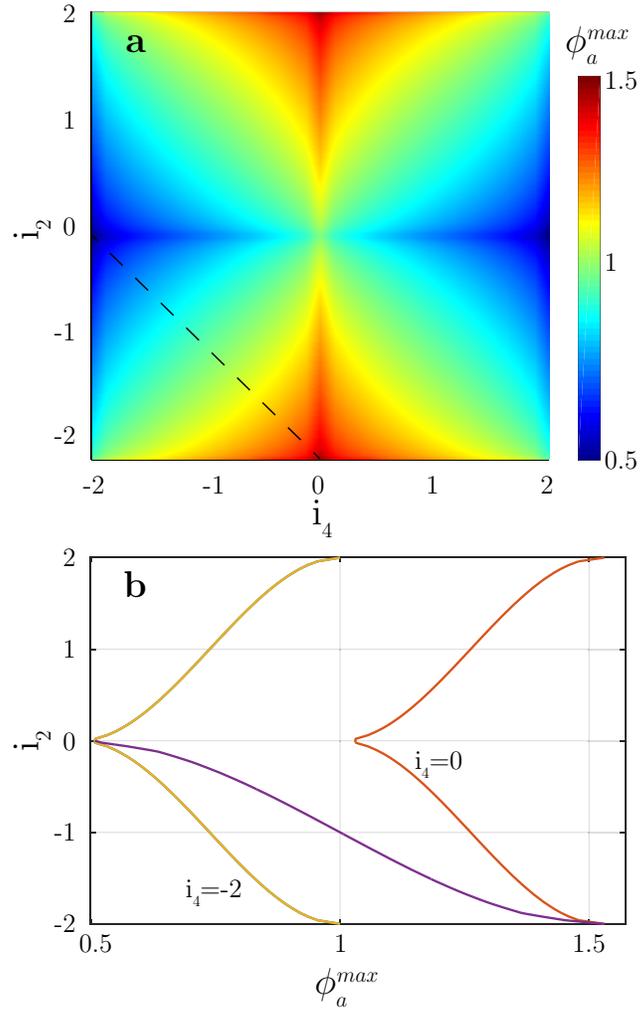}}
  \caption{(a) The surface of $\phi_{a}^{max}(i_4,i_2)$ for $\beta_L=0.1$ and $a_k=1$. (b) Line cuts $\phi_{a}^{max}(i_2)$ at $i_4=-2$ and $0$ and along the diagonal dashed line in (a).}
  \label{fig4}
\end{figure}

Alternatively, the span of $\phi_{a}^{max}$ can be significantly increased by utilizing the two control currents $i_2$ and $i_4$ concurrently. We define $\phi^{max}_{a}(i_4,i_2)$ to be a solution of (\ref{dd}) for $i_2\leq a_2-a_1$ and of (\ref{dd2}) for $i_2> a_2-a_1$, which depends on both $i_2$ and $i_4$. Figure \ref{fig4}(a) presents the two-dimensional surface $\phi_{a}^{max}(i_4,i_2)$ for the case of a symmetric mSQUID. The span of $\phi_{a}^{max}$ along the vertical line cuts at $i_4=-2$ and $i_4=0$ with only $i_2$ varying is displayed in Fig. \ref{fig4}(b), which show that the span equals $1/2$ (for small $\beta_L$). However, by using both control currents the span can be significantly increased, as demonstrated in Fig. \ref{fig4}(a). Thus, along the diagonal dashed line which connects points $(i_4,i_2)=(-2,0)$ and $(0,-2)$, where the function $\phi_{a}^{max}(i_4,i_2)$ attains its minimum and maximum respectively, the span equals $1$ as shown by the purple curve in Fig. \ref{fig4}(b). By increasing the inductance to $\beta_L=1$ the span of $\phi_{a}^{max}(i_4,i_2)$ reaches $1.5$  and hence an electrical tunability of the mSQUID by more than a full flux period can be achieved.

\begin{figure}[c]
\centering
  \makebox[\textwidth]{\hspace*{10cm}\includegraphics[trim=50 520 50 50, clip,scale=1]{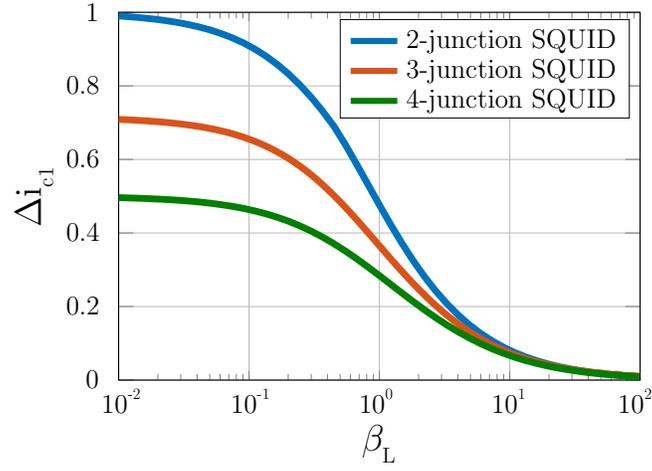}}
  \caption{ The modulation depth of the critical current as a function of $\beta_L$ for conventional 2-junction SQUID and for 3- and 4-junction mSQUIDs with $a_k=1$.}
  \label{fig6}
\end{figure}

An additional important characteristic of the interference pattern is the modulation depth of the critical current, which affects the sensitivity of SQUIDs. We define the modulation depth of the critical current of the mSQUID as $\Delta i_{c1}=(\max i_{c1}-\min i_{c1})/\max i_{c1}$. Figure \ref{fig6} shows the comparison of the modulation depth of a regular 2-junction SQUID with symmetric mSQUIDs as a function of $\beta_L$. In contrast to regular SQUIDs for which $\Delta i_{c}=1$ is attained as $\beta_L \rightarrow 0$, the maximum attainable modulation depth in the 4-junction mSQUID is only $\Delta i_{c1}=0.5$ due to the presence of the additional junctions in the loop. Note, however, that in conventional SQUIDs the optimal sensitivity is usually attained for $\beta_{L}\simeq 1$ for which $\Delta i_{c1}\simeq 0.5$ \cite{TC}. Since this modulation depth can be attained in mSQUID using a lower $\beta_L$ we expect that by proper parameter design the optimal achievable sensitivity of the mSQUIDs should be comparable to that of conventional SQUIDs.

\begin{figure}[!htb]
\centering
  \makebox[\textwidth]{\hspace*{10cm}\includegraphics[trim=50 480 50 150, clip,scale=1]{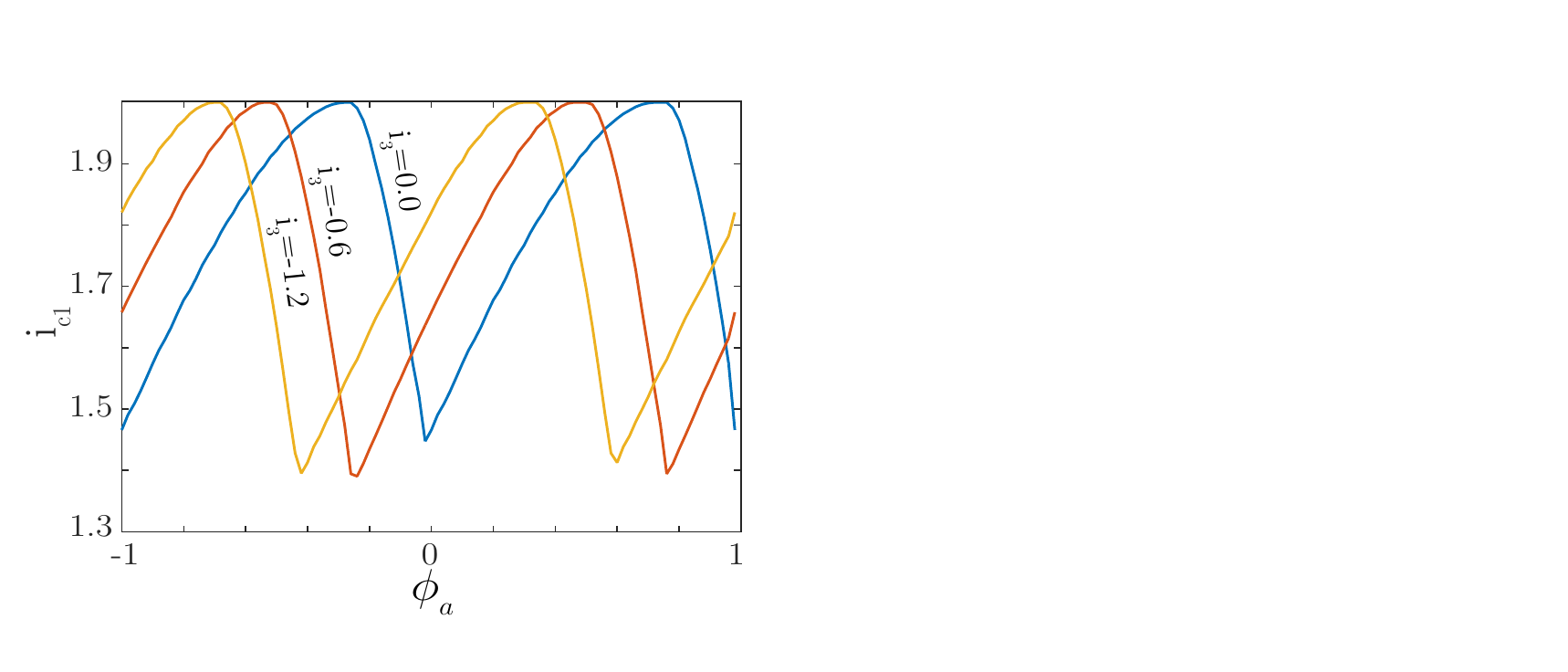}}
  \caption{The interference pattern of a 4-junction mSQUID configured with $i_2$ as a drain current and $i_3$ and $i_4$ as control currents for $\beta_L=1$, $a_k=1$, $i_4=0$, and $i_3=0,\,-0.6,\,-1.2$.}
  \label{fig5}
\end{figure}

So far we have discussed the properties of the mSQUID in a symmetric measurement setup in which $i_3$ serves as the drain terminal. The above derivations can be readily generalized to the case of an asymmetric circuit in which $i_2$ is the drain terminal while $i_3$ and $i_4$ serve as the control currents. It can be shown that the general behavior of the mSQUID in these two schemes is quite similar and has the same modulation depth. The main difference, however, is that the interference patterns in the asymmetric scheme are significantly skewed as demonstrated in Fig. \ref{fig5} due to the  asymmetry between the two arms of the SQUID loop. This configuration may have the advantage of enhanced sensitivity in the steeply varying region due to the considerable increase in the transfer function
 $|di_{c1}/d\phi_a|$. As in the symmetric setup, this region of enhanced sensitivity can be electrically shifted to any value of the applied field using the control currents which allow shifting the interference pattern by a full period when applied concurrently.

\subsection{Determination of current-phase relations}

In our model we assumed a sinusoidal current-phase relation for all junctions. However, we can define an arbitrary current-phase relation as $J=F(\varphi)$. The mSQUID allows a direct measurement of $F$ for each of the junctions as follows. The $\phi_{a}^{max}(i_2)$ curve
for a fixed $i_4$ and $i_2<a_2-a_1$ (white curve in Fig. \ref{Fig2}(a) below the demarcation line) is given, using (\ref{dd}), by
\be
\varphi_2=-\pi/2-\varphi_3-\varphi_4-2\pi\phi_a-\frac{\pi\beta_L}4(a_1+j_2+j_3+j_4), \label{dd3}
\ee
where $\varphi_3$, $\varphi_4$, $j_3$ and $j_4$ are constants, or, in a more compact form, as
\be
\varphi_2 = - 2\pi\phi_a -\pi\beta_L j_2/4 +\mu_1,
\ee
where $\mu_1$ is a constant controlled by $i_4$. As a result, the current flowing through junction 2 is given by
\be\label{newE0}
J_2= F(-2\pi\phi_a -\pi\beta_L j_2/4+\mu_1),
\ee
while the control current $I_2$ is described by
\be \label{newE}
    I_2=J_2-a_1I_0= F(-2\pi\phi_a -\pi\beta_L i_2/4+\mu_2)-a_1I_0. \label{cfrf}
\ee
The relations (\ref{newE0}) and (\ref{newE}) show that the branch of $\phi_{a}^{max}(i_2)$ that lies  below the demarcation line directly describes the current-phase function $F$ of junction 2 upon proper rescaling of the axes. The $J_2$ axis of $J_2=F(\varphi_2)$ is given by the experimentally measured $I_2$ axis of $\phi_{a}^{max}(I_2)$ as follows. The extent of $I_2$ below the demarcation line is equal to the extent of $J_2$ and hence only a translational shift with no rescaling is required. The shift of the $I_2$ axis is readily determined by the fact that the central point of the lower branch of $\phi_{a}^{max}(i_2)$ (point (c) in Fig. \ref{Fig2}(a)) corresponds to $J_2=0$. The $\varphi_2$ axis of $J_2=F(\varphi_2)$ is described by (\ref{newE0}), which in the case of $\beta_L \ll 1$ is given by rescaling the $\phi_{a}$ axis of $\phi_{a}^{max}(I_2)$ by $2\pi$ and a shift $\mu_1$ that can be determined from a self-consistent evaluation of the rest of the junctions. In the general case, $\beta_L$ can be estimated from the modulation depth and then the $\varphi_2$ axis can be rescaled based on (\ref{newE0}) using the experimentally derived values of $J_2$ at the corresponding values of $\phi_{a}$. In a similar manner, by exchanging the role of the terminals the current-phase relations of all the four junctions can be determined independently. This novel property of the mSQUID provides a new tool for the study of current-phase relations in unconventional materials and junctions \cite{Golubov2004,cp1,cp2,cp3,cp4}.

\section{Three-terminal mSQUID}

In this section we analyze the 3-junction, 3-terminal mSQUID configured as shown in Fig. \ref{fig1}(b). Using the same notation as above, we find the critical current $i_{c1}$ as a function of the control current $i_2$ and the externally applied flux $\phi_a$ using the Lagrange multipliers method.

The fluxoid relation for the 3-junction mSQUID is given by
\be
\varphi_1+\varphi_2+\varphi_3 + 2\pi \phi_a+\pi\beta_L j=2\pi n, \label{fr3jj}
\ee
where $j$ is the circulating current
\be
j=(a_1\sin\varphi_1+a_2\sin\varphi_2+a_3\sin\varphi_3)/3.
\ee
Using two side conditions -- the fluxoid relation (\ref{fr3jj}) and the expression for $i_2$ in (\ref{oneof}) -- we write the Lagrangian
\be
\begin{aligned}
\mathcal{L}(\varphi_1,\varphi_2,\varphi_3,\lambda_1,\lambda_2) =& a_1\sin(\varphi_1)-a_4\sin(\varphi_4) \\&+ \lambda_1(\varphi_1+\varphi_2+\varphi_3 +\pi\beta_L j)\\& + \lambda_2(i_2-a_2\sin(\varphi_2)+a_1\sin(\varphi_1)).
\end{aligned}
\ee
The critical points of $\mathcal{L}$, which correspond to a critical state of the mSQUID, satisfy the relation
\be
 a_3 \cos\varphi_{c3}= \frac{-\cos\varphi_{c1} \cos\varphi_{c2}}{r\cos\varphi_{c1}+\frac1 r\cos\varphi_{c2}+\pi\beta_L \cos\varphi_{c1} \cos\varphi_{c2}}.
\ee
The three phases completely define the state of the mSQUID and can be found by assigning the values $(\varphi_{c1},\varphi_{c2})$ on a two-dimensional grid and calculating the phase $\varphi_{c3}$ using (\theequation).

\begin{figure}[!htb]
\centering
  \makebox[\textwidth]{\hspace*{10cm}\includegraphics[trim=50 350 50 50, clip,scale=1,width=\textwidth]{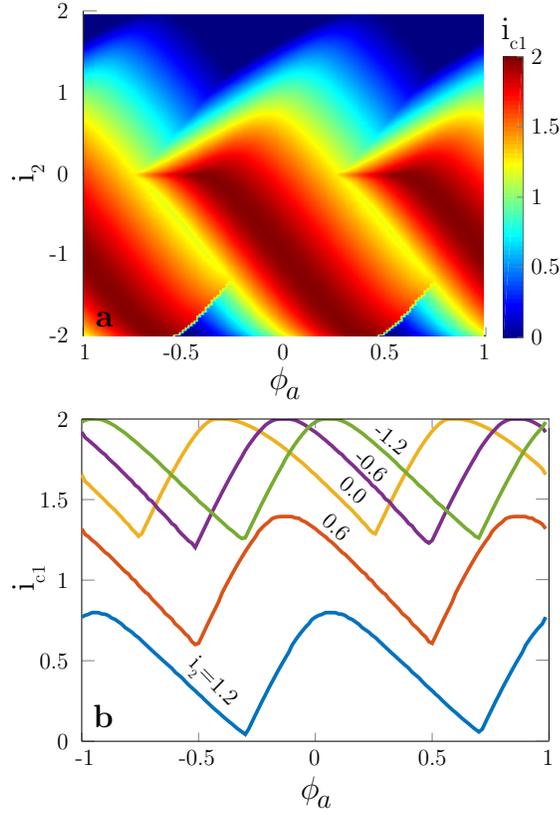}}
  \caption{(a) The critical current $i_{c1}$ of a 3-terminal mSQUID as a function of the bias current $i_2$ and the external flux $\phi_a$ for $\beta_L=1$ and  $a_k=1$. (b) Interference pattern $i_{c1} (\phi_a)$ for the control current $i_2 = -1.2,\,-0.6,\,0,\,0.6$, and $1.2$.}
  \label{fig7}
\end{figure}

An example of the interference pattern as a function of $\phi_a$ and $i_2$ is shown in Fig. \ref{fig7}(a). The structure of the interference pattern for the 3-junction mSQUID is similar to that of the 4-junction mSQUID in Fig. \ref{Fig2}(a). The critical current $i_{c1}$ satisfies $i_{c1} \leq a_1 + a_3$ and the value of $i_2$ satisfies $-(a_2 +a_1) \leq i_2 \leq a_2 +a_1$. The demarcation line of $i_{c1}^{max}$ is located at $i_2=a_2-a_1$. For $i_2 \leq a_2 - a_1$, the interference
pattern $i_c(\phi_a)$ is shifted to the left along the $\phi_a$ axis as $i_2$ increases, while for $i_2 > a_2 - a_1$ it is shifted to the right and its amplitude linearly decreases as in the 4-junction mSQUID. Note that the shape of the interference pattern in the two cases is different. The 3-junction mSQUID has a larger modulation of the critical current as shown in Fig. \ref{fig6} due to fewer junctions in the loop. In addition, the asymmetric structure of the 3-junction mSQUID, with one junction in the left arm and two junctions in the right one, causes a shift in the interference patterns and a skewed structure as shown in Fig. \ref{fig7}(b).

In a 3-junction mSQUID, we can find the curve of the maximum critical current $\phi_{a}^{max}(i_2)$ as follows. When $i_2>a_2-a_1$, we have $\varphi_{c2}=\pi/2$, $\varphi_{c3}=-\pi/2$ and the maximum of the critical current satisfies
\be
\varphi_{c1}=-2\pi\phi_a- \beta_L\pi(a_1\sin\varphi_{c1}+a_2-a_3)/3,\label{te1}
\ee
and when $i_2<a_2-a_1$ we have $\varphi_{c1}=\pi/2$, $\varphi_{c3}=-\pi/2$ and
\be
\varphi_{c2}=-2\pi\phi_a- \beta_L\pi(a_1+a_2\sin\varphi_{c2}-a_3)/3.\label{te2}
\ee
Note that Equations (\ref{te1}) and (\ref{te2}) are invariant under the transformation $G_1$ in (\ref{trtr1}).

\begin{figure}[ht]
\centering
  \makebox[\textwidth]{\includegraphics[trim=50 200 0 150, clip,scale=.9]{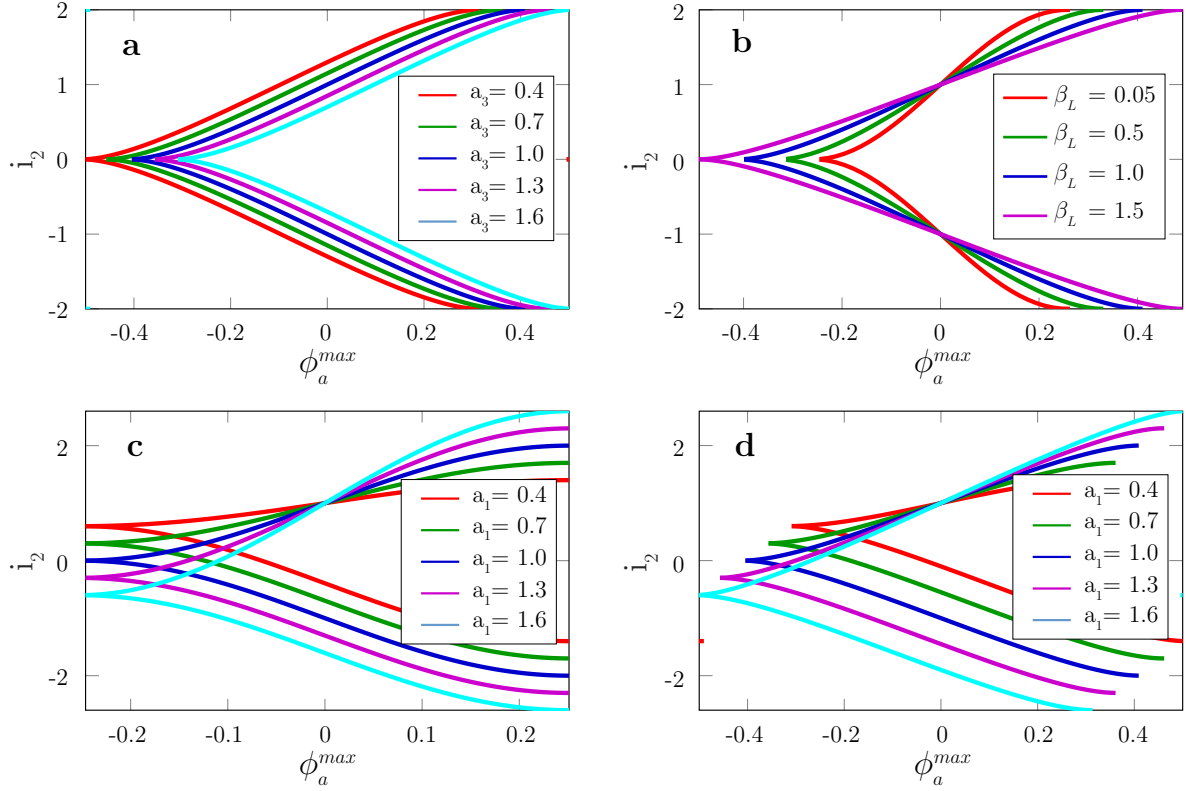}}
  \caption{The function $\phi_{a}^{max}(i_2)$ for the 3-terminal mSQUID, which describes the shift of the interference pattern  as a function of the control current $i_2$ for the following parameter values: (a) $a_1=1$, $a_2=1$, $\beta_L=1$, and $a_3=0.4,\,0.7,\,1,\,1.3,$ $1.6$. (b) $a_k=1$, and $\beta_L=0.05,\,0.5,\,1,$ $1.5$. (c) $a_2=1$, $a_3=1$, $\beta_L=0.05$, and $a_1=0.4,\,0.7,\,1,\,1.3,$ $1.6$. (d) $a_2=1$, $a_3=1$, $\beta_L=1$, and $a_1=0.4,\,0.7,\,1,\,1.3,$ $1.6$.}
  \label{fig8}
\end{figure}

By solving (\ref{te1}) and (\ref{te2}), we find the dependence of $\phi_{a}^{max}(i_2)$ on the various parameters as shown in Fig. \ref{fig8}. Variation of $a_3$ causes a horizontal shift in $\phi_{a}^{max}(i_2)$, as seen in Fig. \ref{fig8}(a), similar to the behavior in Figs. \ref{fig3}(a) and \ref{fig3}(b). The horizontal extent of the $\phi_{a}^{max}(i_2)$ is determined by $\beta_L$, as shown in Fig. \ref{fig8}(b); this is similar to Fig. \ref{fig3}(d), although the relative shift of the curves is different. Also note that the maximum shift of the interference pattern in a 3-junction mSQUID upon varying $i_2$, is smaller than in the 4-junction mSQUID, upon using two control currents concurrently. Figure \ref{fig8}(c) shows that variation of $a_1$ breaks the symmetry between the upper and lower branches of $\phi_{a}^{max}(i_2)$ similar to Fig. \ref{fig3}(c). Note that similarly to the derivation for the 4-junction mSQUID, the horizontal span of the upper branch of $\phi_{a}^{max}(i_2)$ in the 3-junction case is $1/2+\beta_L a_1/3$ and that of the lower branch is $1/2+\beta_L a_2/3$. Therefore, at higher $\beta_L$, the variation of $a_1$ also results in an uneven horizontal span of the upper and the lower branches of $\phi_{a}^{max}(i_2)$ as presented in Fig. \ref{fig8}(d). Finally, transformation $G_1$ dictates that varying $a_2$ instead of $a_1$ will result in reflection of the curves in Figs. \ref{fig8}(c) and \ref{fig8}(d) about the x-axis.

\section{Summary}

We have modeled and analyzed the dc behavior of 3-terminal, 3-junction and 4-terminal, 4-junction mSQUIDs. The extra degrees of freedom in these devices allow a continuous shift of the interference pattern of the critical current with respect to the applied field by applying control currents to the additional terminals. The 3-terminal device has the advantage of larger modulation depth of the critical current, but the interference pattern can be shifted by only about half a period. This is sufficient for attaining maximum sensitivity of the mSQUID over a full operational range of applied fields. The 4-terminal device has a somewhat lower modulation depth but has a number of advantages. Its symmetric structure allowed recent fabrication of a 4-terminal nanoSQUID on the apex of a sharp tip \cite{ExperPaper}, the behavior of which is well described by the presented model. By using two control currents concurrently, the interference pattern of the 4-terminal mSQUID can be readily shifted by a full period. Besides extending the range amenable to  accurate measurement, this tunability makes the device especially suitable for noise reduction using fully electrical  schemes. By connecting the 4-terminal mSQUID to an asymmetric electrical circuit a skewed interference pattern can be obtained, providing additional enhancement of the device sensitivity. Finally, we have shown that the mSQUID provides a new tool for direct measurement of the current-phase relations of the individual Josephson junctions and weak links that can be utilized for study of unconventional superconductors. Multi-terminal, multi-junction nanoSQUIDs are therefore highly promising sensors for nanoscale scanning SQUID microscopy with in-situ electrical tunability that allows operation at maximum sensitivity over a broad range of magnetic fields.

\section*{Acknowledgments }
We would like to thank Martin E. Huber and Alexander Gurevich for useful
discussions and suggestions and Eitan Levin for his contribution during the initial stage of this work.
This research was supported by the US-Israel Binational Science Foundation (BSF grant 2014155), by the Israel Science Foundation (grant No. 132/14), and by Rosa and Emilio Segr\'{e} Research Award.

\bibliographystyle{iopart-num_custom_2}
\bibliography{4jjRefs}
\newpage

\end{document}